\documentclass[11pt, a4paper]{article}

\usepackage[utf8]{inputenc}
\usepackage[T1]{fontenc}
\usepackage{lmodern}
\usepackage{microtype}

\date{}

\usepackage[
  top=20mm, bottom=20mm,
  left=25mm, right=25mm
]{geometry}

\usepackage{amsmath, amssymb, amsthm}
\usepackage{mathtools}
\usepackage{bm}

\usepackage{graphicx}
\usepackage{float}
\usepackage{booktabs}
\usepackage{caption}
\usepackage{subcaption}

\DeclareCaptionLabelFormat{bold-fig}{\textbf{#1~#2}}
\DeclareCaptionTextFormat{sentence-bold}{%
  #1%
}
\usepackage{algorithm}
\usepackage{algpseudocode}

\usepackage[
  colorlinks = true,
  linkcolor  = blue!70!black,
  citecolor  = blue!70!black,
  urlcolor   = blue!70!black
]{hyperref}
\usepackage[capitalize, noabbrev]{cleveref}

\usepackage{xcolor}
\usepackage{enumitem}
\usepackage{lipsum}             

\newcommand{\authmark}[1]{\ensuremath{^{#1}}}
\newcommand{\affilmark}[1]{\ensuremath{^{#1}}}

\title{
  \vspace{-8mm}
  \LARGE\bfseries Interaction mechanics of acoustic cavitation with fibrin networks\\[4pt]
  
}

\author{
  Aarushi Bhargava\authmark{1*,2},\;
  Gaurav Gardi\authmark{2},\;
  Metin Sitti\authmark{2*,3},\;
  \\[6pt]
  \small
  \affilmark{1}Department of Biomedical Engineering, University of Wisconsin-Madison, Madison, WI, 53706.\\
  \small
  \affilmark{2}Physical Intelligence Department, Max Planck Institute for Intelligent Systems, Stuttgart, Germany, 70569.\\\small
  \affilmark{3}School of Medicine and College of Engineering, Koç University, Istanbul, Turkey, 34450\\
  \small
  *Corresponding author(s). E-mail(s):
    \texttt{aarushi.bhargava@wisc.edu};
    \texttt{msitti@ku.edu.tr};
}
\begin{document}

\maketitle
\thispagestyle{empty}

\begin{abstract}
\noindent
Stiff and dense fibrin networks in chronic blood clots impede drug penetration, limiting the efficacy of thrombolytic therapies. Acoustic cavitation of microbubbles is a promising strategy to enhance drug delivery in soft tissues. However, the interaction of these bubbles with stiff fibrin networks has yet to be investigated. Here, we show that ultrasound-driven bubbles undergoing periodic oscillations can penetrate and alter dense fibrin networks. The penetrated bubbles create three-dimensional paths that enable nanobeads (matrix transport markers) to infiltrate up to 200 \textmu m deep into the mesh. Radial bubble oscillation is found to be the dominant forcing mechanism on fibrin fibers. Combining mechanical measurements with these observations reveals that the radial stress from a single bubble oscillation is far below the fracture strength of fibrin fibers. Instead, repeated sub-fracture loading from thousands of oscillations progressively accumulates damage and softens the network until it yields - the plausible mechanism for bubble penetration through the fiber mesh. We further explored this fibrin softening at a range of peak applied forces. At low force, the fibrin network initially softens, but is resistant to further damage after hundreds of cycles. At higher forces, networks continue to soften without reaching a stable state, indicating the progressive accumulation of damage. These results show that cavitation can enhance matrix transport in dense fiber mesh by softening and structurally altering fibrin networks. The underlying physics is governed by the viscoplastic mechanics of bubble-fibrin interactions. These findings establish a mechanistic framework to design comprehensive treatment strategies for fibrotic aged clots.
\end{abstract}


\bigskip
\hrule
\bigskip

\section{Introduction}
\label{sec:intro}

Ultrasound cavitation refers to the dynamic response of gas bubbles to an acoustic pressure field \cite{leighton2012acoustic}. At low acoustic intensities, bubbles undergo small, steady oscillations around their equilibrium radius (stable cavitation), whereas at higher intensities they experience rapid, large-amplitude expansion and collapse (inertial cavitation). Cavitation contributes to a range of biomedical applications, including drug delivery, blood-brain barrier opening, sonoporation, cell and tissue fragmentation, and mechanical stimulation \cite{burgess2015focusedbbb, arrieta2024nueromod, miller2002sonoporation, wright2012vitrothrombo, khokhlova2015histotripsyrev}. The underlying mechanisms can include a combination of microjetting, shock waves, acoustic radiation force, oscillatory force, and microstreaming effects depending on acoustic parameters and surrounding tissue properties \cite{burgess2015focusedbbb, miller2002sonoporation, khokhlova2015histotripsyrev, coussios2008applications}.

Cavitation has also emerged as a promising adjuvant for thrombolysis \cite{wright2012vitrothrombo, datta2006correlationthrombo}. Blood clots (thrombi) underlie many arterial and venous occlusive disorders, including ischemic stroke, myocardial infarction, deep vein thrombosis, and pulmonary embolism \cite{epidemiologyvte2003circ, dvtrisk2014dvt, heartandstroke2017heart, weisel2011coronary}. Clots consist of fibrin, a three-dimensional mesh-like protein network, that entraps red blood cells (RBCs) and platelets \cite{thrombolysistime2023thrombuscomp}. Enhanced drug delivery due to fluid mixing from stable cavitation, and large strains from bubbles undergoing inertial cavitation have both been shown to increase thrombolysis \cite{petit2015fibrin}. However, these strategies have shown limited efficacy in retracted clots, which are characteristic of aged or chronic thrombi \cite{sutton2013clot, hendleydoublingdown2022plos, holland2024histotripsyholl}. Retraction involves substantial structural remodeling where dense and stiff external fibrin layers
develop and enclose a core of tightly packed RBCs and platelets \cite{fibrinfunction2018cardth, rmbolism2021artscl}. Fibrin has distinct mechanical properties from most tissues including high extensibility 
(100\%-150\%), viscoelasticity, strain stiffening behavior under large strains, and Mullins effect \cite{weisel2011coronary, fibrinextraordextensi2006weisel, brown2009multiscale}. These distinct mechanical properties allow retracted clots to withstand large forces and prevent drug penetration into the clot core \cite{alteredfibthrombolysis2002circ, fibronolysistight1996thrandhem}. As a consequence, current cavitation approaches achieve only partial thrombolysis, with fibrin-rich regions remaining intact \cite{hendleydoublingdown2022plos, holland2024histotripsyholl}. These limitations highlight the need for strategies that directly target the fibrin network.

Lipid-coated microbubbles undergoing ultrasound-driven periodic oscillations have recently been explored as a means to mechanically penetrate and alter soft biological materials \cite{wan2015cavitationgail}. However, most prior works have focused on agarose, soft tissues, or soft fibrin hydrogels \cite{bezer2025microbubblebrain, ferrara2009microbubble, acconcia2013interactions}. These materials are relatively soft, have low extensibility, or exhibit weak strain-stiffening, differing substantially from the dense, stiff fibrin networks found in retracted clots. In addition, the mechanisms of microbubble interaction with surrounding materials remain poorly understood. Proposed types of bubble action include microstreaming-induced shear, radiation force, and radial oscillatory forces, yet their relative roles and the material response to these forces are unclear \cite{coussios2008applications}. In this study, we demonstrate that forces from ultrasound-driven microbubbles can disrupt dense, stiff fibrin networks and thereby enhance drug penetration. Through systematic experiments, we identify the physical mechanisms governing bubble-fibrin interactions and delineate the conditions under which fibrin disruption occurs. These insights may inform the design of ultrasound-based strategies aimed at achieving complete thrombolysis.

\section*{Results}
\subsection*{Bubbles can penetrate dense fibrin networks}

To examine microbubble interactions with fibrin networks, we formed hydrogels with fibrin concentrations ranging from 3-10 mg/mL in custom-built acoustically and optically transparent chambers. Varying the concentration produced networks with distinct stiffness and porosity categorized as low (700 Pa, 8-10 \textmu m), medium (1300 Pa, 3-5 \textmu m) and high (1600 Pa, $<$ 1 \textmu m) density networks, Figs. 1A and 1B. These concentrations reflected the variation in characteristics from acute to aged clots \cite{retractedcompos2018jtt, zkabczyk2023porosityclots}. The stiffness was quantified using Atomic Force Microscopy (AFM) in force spectroscopy mode using standard methodology \cite{afmhydrogelproto2021measuring}. A perfusion channel along one edge of the chamber delivered SonoVue\copyright{} microbubbles (0.7-10 \textmu m) and fluorescent nanoparticles ($\sim$200 nm), with inlet and outlet ports enabling continuous replenishment. The chamber was placed in a water tank under a microscope, and a 1 MHz ultrasound transducer was positioned with its face perpendicular to the channel, Fig. S1. High-speed imaging at 44,000 frames per second captured bubble dynamics during insonation. Fluorescent nanoparticles, used to estimate macromolecular transport through the matrix, allowed visualization of fibrin disruption and penetration.

Microbubbles were observed to undergo radial oscillations and move along the direction of the ultrasound pulse (1 ms pulse duration, 150 Hz pulse repetition frequency) to come in contact with the boundary between fluid and fibrin networks. Bubbles then penetrated the fibrin networks across all tested acoustic pressures (0.3 - 0.5 MPa), Movies S1-S3, and in all tested densities of fibrin networks, Movies S2, S4, and S5. Figure 1C (and Movie S2) shows a representative example in which two bubbles enter a high-density fibrin mesh at 0.4 MPa and continue traveling several micrometers into the network over four successive pulses. We quantified the motion of the microbubbles inside fibrin networks using a custom-built algorithm for bubble detection and tracking (details in Methods). As expected, the bubble speed decreased with increasing density and stiffness of the fiber mesh, Fig. 1D, due to greater mechanical resistance in denser networks. In contrast, bubble speed showed no clear dependence on acoustic pressure within the tested range in high-density mesh, Fig. 1E. The total distance traveled by each bubble within a single pulse (composed of 1000 pressure cycles) also showed similar trends with increasing fibrin density and acoustic pressure, Figs. S2(A) and S2(B). At the highest pressure tested (0.5 MPa), bubbles often traversed the imaging field in $<$ 1000 cycles, leading to truncated distance measurements constrained by the imaging window.

\begin{figure}[p]
  \centering
  \includegraphics[width=.7\textwidth, height=0.8\textheight, keepaspectratio]{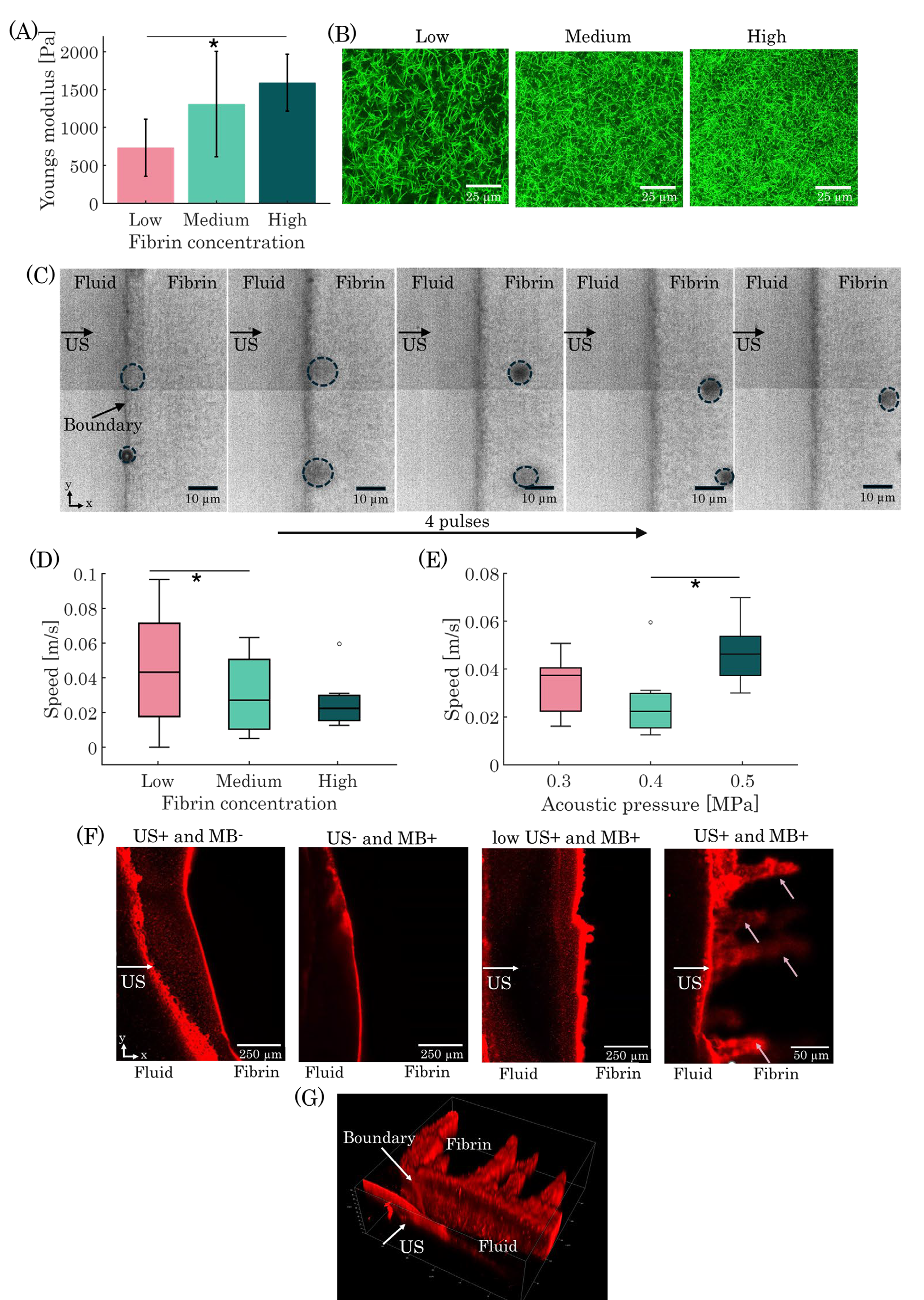}
  \caption{\textbf{Interaction of ultrasound-driven bubbles with fibrin networks.}
    Interaction of ultrasound-driven bubbles with fibrin networks. Young's modulus (A) and fibrin architecture (B) of the three densities of fibrin hydrogels tested - low, medium, and high. The density is varied by varying the fibrin concentration. The scale bar in (B) is 25 \textmu m. (C) Penetration of bubbles from fluid (left) into high-density fibrin hydrogel (right) at 0.4 MPa of ultrasound pressure (wave direction from left to right) at different time points, as they move across the imaging window, Movie S2. Bubbles are highlighted with circles of dashed lines. The penetration occurs over 4 pulses (4000 cycles) at a pulse repetition frequency of 150 Hz. The bubbles appear blurry when they move out of the focal plane of the camera implying three-dimensional motion of bubbles in the mesh. The scale bar is 10 \textmu m. Maximum speed of bubbles inside the mesh of (D) different densities at 0.4 MPa and (E) across various acoustic pressures in high-density mesh, respectively. The acoustic pressure was kept below the inertial cavitation threshold. (F) Nanobead penetration under various ultrasound (US) conditions. No nanobead penetration is seen in the first three cases when there are no microbubbles (US+ and MB-) - Movie S6, when there is no ultrasound (US- and MB+) - Movie S7, and when microbubbles are present but do not penetrate at low (0.01 MPa) ultrasound pressures (low US+ and MB+) - Movie S8. Representative case (US+ and MB+) showing nanoparticle penetration inside fibrin mesh at 0.5 MPa, Movie S9. The arrows highlight selective nanoparticle penetration only along paths created by bubbles. The scale bars are different and highlighted in the images for all cases. (G) Three-dimensional representation of the US+ and MB+ case at 0.5 MPa in high-density mesh, highlighting that bubble penetration occurs along all three axes. Up to 200 \textmu m penetration is observed along the ultrasound wave direction. The asterisk symbol in all figures indicates statistical significance between groups.}
  \label{fig:1}
\end{figure}

\subsection{Nanoparticle infiltration occurs along the paths created by bubbles}
\label{sec:results:Nanoparticle infiltration occurs along the paths created by bubbles}

Nanoparticle infiltration into fibrin networks occurred exclusively when microbubbles were actively driven by ultrasound at $\geq0.3$ MPa (MB+, US+ condition), shown in Fig. 1F (right) and Movie S9. Nanobeads were confined to discrete tracks formed by penetrating bubbles, indicating that transport enhancement was spatially localized along bubble-generated paths rather than uniformly throughout the network. Confocal z-stack imaging revealed that these paths extended in three dimensions, reaching up to $\sim$200 \textmu m along the direction of acoustic wave propagation, Fig. 1G. The measured tunnel widths ($\sim$19.6 \textmu m) substantially exceeded the intrinsic fibrin pore size ($<$ 1 \textmu m), confirming significant local fiber displacement or rupture induced by bubble motion.

Although the nanoparticle diameter is smaller than the nominal pore size of the fibrin networks tested, passive infiltration does not occur spontaneously. The heterogeneous network architecture, tortuous transport pathways, reduced permeability, and elevated interstitial pressures collectively impede transport in the absence of strong external forcing or chemical degradation, consistent with observations in dense blood clots \cite{wufsus2013clotpermeability, shibeko2020redistributiontpaflux}. Three control conditions confirm that both microbubbles and sufficient acoustic driving are necessary prerequisites for this transport enhancement. When ultrasound was applied without microbubbles (US+ and MB-, 0.5 MPa), no nanoparticle infiltration was observed, Fig. 1F and Movie S6, demonstrating that ultrasound-induced shear alone cannot disrupt the fibrin mesh. When microbubbles were present without ultrasound (US- and MB+), infiltration was equally absent, Movie S7, confirming that bubbles require acoustic activation for penetration. When ultrasound was applied at 0.01 MPa in the presence of microbubbles (low US+ and MB+), the acoustic pressure was insufficient to drive bubble penetration into the network, and no nanoparticle transport was observed, Movie S8. Together, these observations establish that (i) ultrasound-activated microbubbles are required to disrupt the fibrin mesh, not ultrasound or microbubbles in isolation. (ii) Effective microbubble-mediated penetration occurs only above a threshold acoustic pressure that lies between 0.01 MPa and 0.3 MPa. More experiments are needed within this range to determine this threshold pressure, and remain a part of future work.

It is important to note that the 200 nm nanobeads used here are substantially larger than most thrombolytic agents ($<$ 20 nm) \cite{mican2019structuralproteinsizedrug}. Therefore, nanobeads serve as only qualitative markers of structural disruption and altered matrix accessibility rather than quantitative predictors of drug transport. Their infiltration along bubble-generated tunnels provides proof that bubble-mediated matrix disruption creates accessible pathways within the fibrin network.

\subsection{Forces applied by the bubbles to penetrate fiber mesh are lower than fiber fracture strength}
\label{sec:results:Forces applied by the bubbles to penetrate fiber mesh are lower than fiber fracture strength}

A plausible mechanism for bubble penetration is that bubble-generated forces break fibrin fibers and lower the resistance to move deeper into the mesh.  To test this, we compare the bubble-generated forces with the fracture strength of fibrin fibers. Fibrin is a unique material which shows viscoelasticity, high extensibility, and strain-stiffening behavior \cite{weisel2011coronary, brown2009multiscale, fibrinextraordextensi2006weisel}. To characterize the mechanical behavior of the fibrin networks used in our experiments, Figs. 1A and 1B, a flat punch indenter was used to apply force on a sample at a rate of 80 \textmu m/s (details in Methods). Figure 2A shows the representative mechanical response of the tested fibrin hydrogels. As the indenter displacement into the material increases, a non-linear increase in stress occurs, exhibiting strain-stiffening. At a threshold displacement, a sawtooth-like response occurs with repeated stress drops, leading to final failure. Consistent with the hierarchical and heterogeneous structure of fibrin, these stress drops reflect sequential fiber failure \cite{boots2022sawtoothrupture, montanari2023puncturingrupture2, piechocka2010fibrinhierarchy}. We defined the stress at which the first drop occurs as the fracture strength. As expected, the fracture strength increases with increasing density of fibrin network, from 0.37 MPa in low-density gels to 0.96 MPa and 1.2 MPa in medium- and high-density networks, respectively.

\begin{figure*}[t!]
\centering
\includegraphics[scale=0.45]{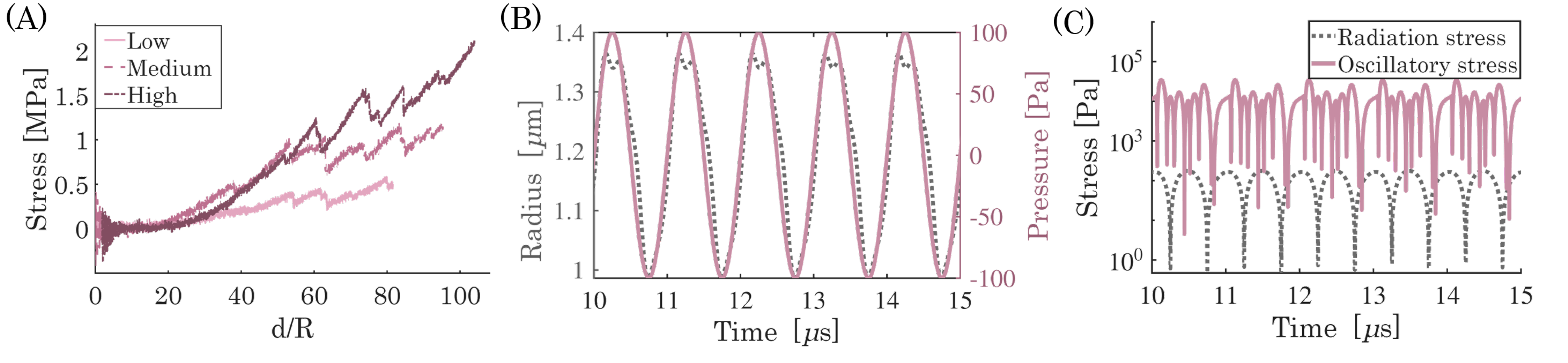}
\caption{(A) Stress response of fibrin hydrogels of low, medium, and high-density during microindentation at 80 \textmu m/s. The x-axis has indenter displacement $d$ normalized with its radius $R$. (B) Model-estimated radial response of a bubble (left) in time domain to an acoustic pressure of 100 kPa (right) at 1 MHz. The equilibrium radius of the bubble is 1.25 \textmu m. (C) Induced stresses from oscillatory and radiation forces of the bubble in the time domain at 1 MHz.}
\label{fig:frog2}
\end{figure*}

The bubble-fibrin interactions in Fig. 1 occur at MHz frequency, at which the radial deformation rate is 10\textsuperscript{4} times higher than the indentation rate used in Fig. 2A. Therefore, a relevant comparison must be made at the MHz deformation and strain rate. Because fibrin is viscoelastic, its fracture strength and elastic modulus increase with strain rate \cite{liu2024fibrinfatigueandvariablerateloading}. Prior studies suggest that fibrin fracture strength scales weakly with strain rate, approximately following a power-law relationship with an exponent of $\sim$0.17 \cite{mooney2006indentationrate, liu2024fibrinfatigueandvariablerateloading,litvinovstrengthfibrin, kliuchnikov2025strengthdamage}. This exponent is derived from a limited set of fracture measurements at relatively low strain rates, since direct mechanical measurements at the bubble-relevant strain rates are not feasible with current equipment capabilities. A conservative extrapolation of this scaling law places the fracture strength at MHz deformation rates to $\sim$6 MPa, exceeding the value measured at a low rate in Fig. 2A.

We next estimated the stresses applied by microbubbles during ultrasound exposure. A modified Rayleigh-Plesset model for lipid-coated bubbles, widely used to predict bubble behavior \cite{marmottant2005model}, was used to compute the two dominant forces expected during stable cavitation- acoustic radiation forces and oscillatory forces. Radiation forces, exerted on a bubble by a non-zero pressure gradient, result in bubble translation. As a bubble translates, it exerts a force on the material along the pressure gradient \cite{lee1993acousticrad}. Oscillatory forces are applied by the volumetric bubble pulsations under ultrasound \cite{jimenez2005bubbleosci}. Microstreaming-induced shear is neglected because it is a secondary force with its magnitude 1-2 orders of magnitude lower than these primary forces \cite{elder1959cavitationstreaming}. As a first comparison with fracture strength, we modeled bubble expansion and associated stresses at 1 MHz and 100 kPa in the free-field, Figures 2B and 2C. Even though radial expansion is maximum in the free-field, the stresses from radiation and oscillatory forces remain below the inferred high-rate fracture strength, and even below the fracture strength measured at 80 \textmu m/s indentation (Fig. 2A).

Bubble radial expansion within fibrin is further suppressed relative to the free field because the stiffness of the surrounding network resists bubble expansion \cite{vlaisavljevich2014effectsstiffnesshisto, Highspeeddeformationinertial2023franck}. The free-field expansion and stress estimates, therefore, represent an upper bound on what bubbles can generate inside the network. In the absence of sufficient quantitative data on fibrin stiffness at high strain rates, it is not possible to predict the precise reduction in radial bubble expansion. Thus, we consider a conservative worst-case scenario in which bubble expansion within the fibrin network matches that in the free field. 

To estimate the corresponding network stress, we constructed a power-law relationship from the mechanical response of fibrin measured at 80 \textmu m/s (Fig. S5) together with its previously reported rate-dependence \cite{mooney2006indentationrate, liu2024fibrinfatigueandvariablerateloading}. The stress on the fibrin network during indentation, $\hat{S}$, is estimated to scale as $\hat{S}= 4500 \bar{d}^{2}\dot{\bar{d}}^{0.22}$, where $\bar{d} = d/R$ is normalized displacement,  $\dot{\bar{d}}=\dot{d}/R$ is normalized displacement rate, $R$ is the indenter radius, and a prefactor of 4500 (unit Pa) is obtained by curve-fitting. For a representative case of a bubble expanding to 1.36 \textmu m (Fig. 2B) in free-field at 100 kPa, the induced displacement is 0.11 \textmu m and the resulting stress within the fibrin network is approximately 41 Pa. Similarly, for a bubble expanding to 4.4 \textmu m at 400 kPa, the resulting stress is approximately 0.08 MPa, two orders of magnitude below the estimated fracture strength at the corresponding rate ($\sim$6 MPa), and well below the conservative free-field comparison above. 

Together, these analyses, under conservative assumptions that overestimate bubble expansion, indicate that the stresses generated during a single oscillation cycle are consistently insufficient to fracture fibrin fibers. Single-cycle rupture is therefore a highly unlikely mechanism of bubble penetration, which raises a central mechanistic question:  How do microbubbles penetrate and travel nearly 200 \textmu m when the forces generated in a single oscillation cycle are insufficient to break individual fibers?

\subsection{Cyclic forces induce fatigue in fibrin networks}
\label{sec:results:Cyclic forces induce fatigue in fibrin networks}

Experiments in Fig. 1 show that bubbles undergo repeated radial oscillations during each ultrasound pulse, applying cyclic forces to the surrounding fibrin network. Therefore, we analyzed the fibrin response over repeated loading cycles using AFM. A spherical bead of 10 \textmu m, which is in the range of expanded bubble size (more details in Methods), was used to indent the hydrogels. It was assumed that the contact area of the bead is similar to half the bubble surface area. Repeated indentations from the bead were performed to mimic the compressive forcing on the material from cyclic expansion-contraction phases of the bubble half-surface. Based on Fig. 2 observations that bubble-induced stresses are below the fibrin fracture strength, the indenter applied the maximum peak force on the gel in the sub-fracture regime, to isolate the effect of low-stress cyclic loading on fibrin networks.

\begin{figure*}[t!]
\centering
\includegraphics[scale=0.8]{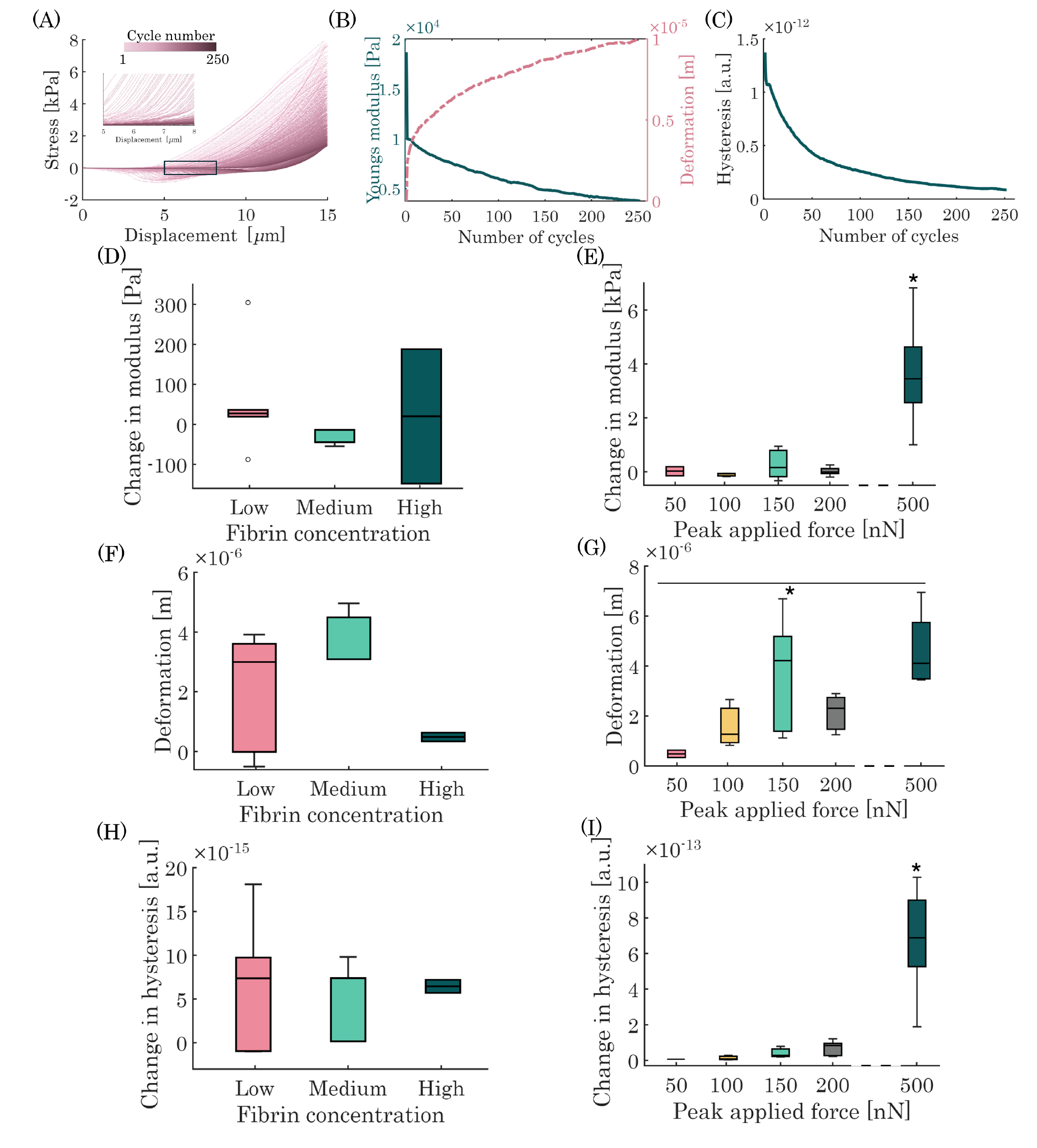}
\caption{Atomic force microscopy measurements to evaluate the response of fibrin networks to repeated stresses lower than fracture strength. (A) Representative example for stress response evolution in high-density fibrin mesh over 250 cycles of loading-unloading. The indenter indents 15 \textmu m into the hydrogel in each cycle with a peak force of 500 nN at 80 \textmu m/s. The inset shows the change in contact point (position at which stress first becomes non-zero during loading) with increasing cycles. (B) Corresponding evolution of Young's modulus and deformation over 250 cycles. Young's modulus is calculated from the initial slope of the unloading curve of a cycle. Deformation corresponds to the difference between the contact point of each cycle and the first cycle, indicating the accumulating plastic deformation in the material. (C) Corresponding hysteresis evolution over successive cycles. Hysteresis is calculated as the area enclosed within the loading-unloading curves of a cycle. Total change in modulus, deformation, and hysteresis in different densities of fibrin networks (D, F, and H) at 50 nN peak applied force, and across various peak applied forces of the indenter (E, G, and I) in high-density mesh, respectively. Change is defined as the difference in value of the metric in the first cycle and the last cycle. The asterisk symbol denotes statistical significance of fibrin networks exposed to 500 nN peak force over all other cases for modulus and hysteresis change, and significance over 50 nN case for deformation. }
\label{fig: Measurements from Atomic Force Microscopy.}
\end{figure*}

It is important to acknowledge a fundamental constraint of the AFM experiments: the loading rates achievable with AFM are orders of magnitude lower than the MHz strain rate experienced by fibrin fibers during bubble oscillations. Moreover, to our knowledge, no published data characterize the fatigue behavior of fibrin networks at such high strain rates, making it infeasible to derive a rate-scaling law of the fatigue response. Such limitations in current experimental techniques represent an important open challenge for the field. The AFM experiments presented here are therefore not intended to replicate the exact loading conditions of bubble oscillations, but rather to interrogate the mechanistic principle: whether sub-fracture cyclic loading produces cumulative and irreversible damage in fibrin networks. This addresses how repeated sub-fracture bubble stresses affect the surrounding fibrin fibers. 

Figure 3A shows the stress-strain response of high-density fibrin hydrogel over 250 cycles at 500 nN peak applied force at a rate of 80 \textmu m/s, consistent with the rate in Fig. 2A. The stress response of the fibrin network during a single loading-unloading cycle in Fig. 3A is nonlinear due to the strain-stiffening effect. Hysteresis also occurs with every cycle, reflecting dissipative losses in the network. Across successive cycles, the strain at which stiffening occurs shifts to larger values. Hysteresis and nonlinearity in the stress decrease with increasing cycle number. For the initial 10 cycles, a significant decrease in peak stress (14.5\%) is observed (Fig. 3B). The peak stress continues to decrease after hundreds of cycles, albeit at a lower rate. These trends at sub-failure loading conditions indicate accumulating damage with weakening of the fibrin networks, characteristic of fatigue phenomena \cite{bai2019fatiguehydrogel}. In contrast to the behavior at high applied forces, damage did not accumulate in high-density networks at low applied forces ($\sim$50 nN), Fig. S3. Instead, the response stabilized after a few hundred cycles, indicating plastic stabilization and network re-arrangement. This is typical of the shakedown effect, which has been reported in many polymers where the fatigue threshold is above the peak stress amplitude \cite{liu2025molecularshakedown, liu2024fibrinfatigueandvariablerateloading, chinh2008shakedown}.

Figure 3B quantifies the deformation at which the stress first becomes non-zero for each cycle during loading. The successive shift in this deformation to higher values, along with residual strain, indicates accumulating permanent deformation with successive cycles. Notably, the duration of each loading-unloading cycle of the probe is in the order of the fast relaxation time of fibrin fibers ($\sim$ms) \cite{liu2010mechanicalgutholdsiglefibers, chinh2008shakedown, liu2024fibrinfatigueandvariablerateloading}, allowing the network to only partially recover before the next cycle. Incomplete recovery allows irreversible changes to accumulate over time. Such changes reflect molecular- and fiber-scale processes including irreversible protein unfolding, crosslink breakage, fiber lengthening, and rupture of stiffer load-bearing fibers \cite{liu2010mechanicalgutholdsiglefibers, fibrinextraordextensi2006weisel, litvinov2017fibrinorigin}. These irreversible changes contribute to dissipative losses, Fig. 3C. Figure 3B also shows the decrease in modulus with successive cycles. As an increasing number of plastically elongated or damaged fibers lose their load-bearing function, the remaining network supports a smaller fraction of the applied load, resulting in a lower modulus. 

The trends in the deformation, modulus change, and hysteresis are not linear. Figure 3B shows that the plastic deformation first increases rapidly and then grows gradually with cycles. From a fiber-scale perspective, this is anticipated as early cycles predominantly engage protofibrils and crosslinks most susceptible to unfolding and bond rupture \cite{liu2024fibrinfatigueandvariablerateloading, chinh2008shakedown, ramanujam2023biomechanicsenergeticsfibrin}. Later cycles involve a progressively smaller population of vulnerable structural elements. This also explains the similar trend observed in modulus and hysteresis for successive cycles, Figs. 3B and 3C. Overall, the fibrin network accumulated approximately 10 \textmu m of permanent extension over 250 cycles and exhibited a decrease of 56\% in modulus. These observations demonstrate substantial softening and progressive weakening of the fibrin mesh under sub-failure cyclic loading.

Next, we analyzed the effect of density and peak applied force on the fatigue response of fibrin networks. Figures 3D, 3F, and 3H show the change in Young's modulus, permanent deformation, and hysteresis, respectively, over 4000 cycles under 50 nN peak applied force for different fibrin densities. Although there was an observable change in these metrics for each density, no statistically significant differences were observed between different densities. A possible reason is that the peak applied force and the corresponding stress were low and below the threshold required for damage accumulation. The initial cycles show changes in these metrics due to network rearrangement. However, the network is able to resist further structural changes and becomes stable, consistent with the shakedown effect rather than fatigue, Fig. S3 \cite{chinh2008shakedown}. In contrast, a different trend was observed when the peak applied force was between 50 - 500 nN in high-density fibrin networks over 4000 cycles. As expected, hysteresis loss, permanent deformation, and the decrease in modulus over multiple cycles increased with increasing applied force. There was statistical significance in all three metrics, particularly for the highest value of applied force (500 nN) tested. Over multiple cycles, the modulus decreased by 50\%, permanent deformation increased by 40\%, and total hysteresis loss was 40\% of the values observed for the first cycle at 500 nN. These observations indicate that microscale fatigue in fibrin gels is initiated only after a force and a corresponding stress threshold.  Overall, these results demonstrate that repeated sub-failure loading induces cumulative and irreversible microstructural damage, weakening the fibrin networks via localized fatigue \cite{bai2019fatiguehydrogel}. Such progressive weakening under cyclic stress provides a mechanistic explanation for how microbubbles, despite exerting forces far below the fracture threshold, can penetrate fibrin networks and create extended paths for potential drug transport during ultrasound exposure.

\section*{Discussion}
In this work, we extend prior studies that demonstrate bubble-tissue interactions \cite{ferrara2009microbubble, acconcia2013interactions, bezer2025microbubblebrain}. We show that bubbles undergoing periodic oscillations under ultrasound can penetrate even dense and stiff fibrin networks. These penetrating bubbles create channels within the mesh, as evidenced by nanobead infiltration (an estimate of matrix transport) up to 200 \textmu m. Modeling and mechanical testing reveal that while oscillatory forces dominate bubble-matrix interactions, the stresses generated during a single oscillation cycle are below the fibrin fracture strength. This discrepancy raises a fundamental mechanistic question: How do microbubbles traverse fibrin networks when bubble forces are insufficient to rupture fibers? We investigate this using AFM. While our AFM indentation experiments do not replicate the loading magnitude and rates of bubble oscillations, they isolate the mechanistic principle. These experiments revealed that repeated sub-failure forcing induces cumulative damage, evident from an increase in permanent deformation and hysteresis, and a decrease in modulus, due to fatigue in fibrin networks. We propose that fatigue is the plausible mechanism leveraged by bubbles for penetration into fibrin networks, where cyclic bubble oscillatory stresses irreversibly deform and weaken fibrin networks, enabling the bubble to move into the fiber mesh.

Bubble infiltration occurred across all fibrin densities and acoustic pressures tested. However, the number of penetrated bubbles depended on the applied ultrasound pressure and the structural properties of fibrin networks. As expected, a higher percentage of bubbles infiltrated the mesh at higher acoustic pressures and in less stiff meshes. Although bubble size is known to influence penetration through controlled single-bubble studies \cite{jones2006trappingstride, beekers2022internalizationkooiman}, our experiments did not maintain strict control of bubble size to reflect in vivo conditions, where polydisperse bubble populations interact with tissue. Bubble coalescence also contributed to the change in the size of penetrating bubbles in our experiments. Coalescence results from attractive secondary Bjerknes forces, which increase with concentration and pressure, especially for bubbles at or below resonance (2.5 \textmu m diameter bubbles have a resonance frequency of 4.6 MHz) \cite{postema2004ultrasound}. Because insonation was performed at 1 MHz, coalescence probability was elevated in our experiments, particularly at higher pressures. Coalescence also altered bubble trajectory, velocity, and maximum penetration distance, as evident in Movie S10. These dynamic changes likely contributed to the non-monotonic velocity and distance trends observed with increasing pressure (Figs. 1E and S2(B)). While coalescence-induced changes were observed, we did not quantify their frequency or explicit contribution to penetration.

Our results indicate that repeated forcing from radial oscillations is the primary driver of fibrin network alteration. Radiation forces and streaming-induced shear may also contribute, but their magnitudes remain far below those of oscillatory forces, Fig. 2C \cite{elder1959cavitationstreaming}. Forces from microjets of collapsing bubbles have also been observed to induce large forces on soft-tissue boundary \cite{brujan2001dynamicsjetting}. While our imaging capabilities limited us from observing microjets, we do not attribute bubble penetration in fibrin to jetting for two reasons: 1. Jetting typically occurs during inertial cavitation of the bubble. Our driving acoustic pressures are below the inertial cavitation threshold of the Sonovue microbubbles ($\sim$ 0.6 MPa in free field) \cite{lin2017sonovuecavitation}. Moreover, the strain stiffening of fibrin will further shift this threshold to higher pressures and reduce bubble expansion \cite{vlaisavljevich2014effectsstiffnesshisto}. 2. Jetting requires asymmetric collapse \cite{brujan2001dynamicsjetting}. While jetting may play a role in initially disrupting the fluid-fibrin boundary interface, it is less likely to occur when the bubbles are fully embedded within fibrin mesh, symmetrically surrounded by material.

Several works have demonstrated that under high-cycle fatigue, sub-fracture loads produce progressive damage, residual deformation, and eventual failure of fibrin networks \cite{liu2024fibrinfatigueandvariablerateloading, ramanujam2023biomechanicsenergeticsfibrin, liu2025molecularshakedown}. Similar trends emerged in our experiments, Fig. 3. We propose that bubbles exploit the same fatigue mechanism to penetrate fibrin networks while applying stresses below the fibrin fracture threshold in a single cycle. This is supported by observations of bubble penetration velocities. At 0.4 MPa acoustic pressure in high-density mesh, the mean bubble velocity is 0.025 m/s, Fig. 1E. To move a distance equal to the maximum pore size of 1 \textmu m at this speed, the bubble will undergo 40 oscillations at 1 MHz. Thus, surrounding fibrin fibers experience repeated cyclic forcing per micrometer of bubble advance. This supports a fatigue-driven mechanism rather than instantaneous fiber rupture. Many studies have observed bubble tunneling in tissue phantoms that are primarily made of agarose or similar materials \cite{ferrara2009microbubble, bezer2025microbubblebrain, vince2023highspeedmicrospheretransport}. Agarose is a viscoelastic material with brittle-like fracture properties and low strain-stiffening behavior \cite{bertula2019strainstiffeiningagarose}. In contrast, fibrin shows high extensibility before fracture, pronounced strain-stiffening, and a hierarchical molecular architecture. Due to these structural differences, the mechanisms of bubble penetration explored in prior works cannot be extrapolated to fibrin-based materials.

In conclusion, our findings highlight a strategy for mechanically enhancing fluid transport and consequent drug delivery in dense and stiff fibrous networks, independent of the drug type. The fibrin response to high-cycle loading reported here can inform the development of computational models that capture fatigue in such hierarchical biopolymer networks, which are currently lacking. Since bubble-fiber mechanical interactions are used to enhance matrix transport, this strategy may extend to other physiological matrices such as collagen and actin \cite{munster2013strainhistory}, which exhibit similar mechanical properties as fibrin. Indeed, several studies have highlighted the rate-dependent high-cycle fatigue under sub-fracture loads in collagen networks, leading to damage accumulation and structural alterations \cite{chawla2024henalcollagenfatigue, vazquez2019cartilagefatigue}. Consequently, the principles demonstrated here may inform therapeutic strategies in pathologies characterized by dense or fibrotic extracellular matrices, including thrombosis, fibrosis, and solid tumors. 

\section*{Methods}

\subsection*{Chamber preparation}
A 1.5 cm$\times$1.5 cm$\times$1 mm chamber was prepared using PDMS. The side walls were composed of thick PDMS spacers while the top and bottom surfaces were made of PDMS sheets (thickness of $\sim$ 200 \textmu m). The chamber was sealed to prevent any leakage or water influx when immersed in the water tank. The chamber featured two side ports with inlet and outlet tubing. This tubing was used to maintain a flow of fresh microbubbles for each experiment. This fluid channel with bubbles was formed along one edge of the chamber, as shown in Fig. S1. The remaining portion of the chamber was occupied by fibrin gel. For enabling gel formation and adhesion, chambers were cleaned and ozone-treated before introducing the gel solution. Fiducial markers were placed at different locations on the bottom surface of the chamber (away from the lens) to assist with high-speed and confocal imaging.

\subsection*{Polymerization of fibrin networks}
Fibrin clots were formed using fibrinogen (Sigma Aldrich), Factor XIII (Coachrom Diagnostica), and thrombin (Sigma Aldrich). Fibrinogen was reconstituted in Tris-HCl buffer comprising 50 mM Tris and 150 mM NaCl, with pH adjusted to 7.4 using HCl. Factor XIII was added to the fibrinogen solution at 20 LU/mL. For making fluorescent fibrin gels, AlexaFluor-488 tagged fibrinogen (Thermo Fisher Scientific) was mixed with native fibrinogen in the buffer solution at a ratio of 1:10. Thrombin was prepared according to the manufacturer's instructions, and diluted in Tris-HCl buffer containing 20 mM CaCl\textsubscript{2}, to a final concentration of 0.7 U/mL. Thrombin and fibrinogen solutions were mixed, poured into the chambers to the height of the inlet/outlet ports, and allowed to polymerize at room temperature for 2 hours. Final fibrinogen concentrations of 3 mg/mL, 6 mg/mL, and 10 mg/mL were used to make low, medium, and high-density gels, respectively. The hydrogels were used within 3 hours after formation for experiments.

\subsection*{Microbubble handling}
Vials of Sonovue (Bracco) were stored at 4\textdegree C and allowed to warm to room temperature for an hour before use in experiments. Sonovue was activated according to the manufacturer's instructions. Briefly, the contents of the vial were reconstituted with PBS and shaken for at least 30 seconds before use. The solution was further diluted in PBS to a final concentration of 10\textsuperscript{7} bubbles/mL, a typical dosage used for drug delivery \cite{petit2015fibrin, coussios2008applications}. Sonovue microbubbles have a lipid monolayer shell enclosing SF\textsubscript{6} gas core. The microbubble diameter range from 0.7 \textmu m to 10 \textmu m with the highest percentage of bubbles falling between 0.8-2.5 \textmu m by number \cite{sennoga2012sonovuesize}. Fluorescent nanobeads (200 nm, excitation/emission: 580/605, Thermo Fisher Scientific) were also additionally added to the microbubble solution. These beads served as markers to assess the matrix's transport properties post cavitation.

\subsection*{Transducer characterization}
Experiments were conducted to measure the pressure profile of the transducer (H269, focal length: 15mm, Sonic Concepts) used in this study. The transducer, immersed in a water tank, was connected to a function generator (AFG 1022, Tektronix Inc.) through an amplifier (150A100C, Amplifier Research). A Precision Acoustics 0.4 mm needle hydrophone (HNA-0400, Onda) was used to acquire pressure measurements. The hydrophone was connected to an oscilloscope (MDO3024, Tektronix Inc.) to visualize and capture the pressure signal. The hydrophone was mounted on a motorized positioning system (LTS300/M, Thorlabs) with its needle pointing towards the transducer, to capture the spatial focal pressure profile in axial and radial directions of the transducer, Fig. S4, at a fixed pressure magnitude. The -6 dB axial and transverse focal distances were 10 mm and 2.5 mm, respectively.

\subsection*{Experimental setup}
A custom-built water tank was designed and placed on the motorized stage of an upright microscope (Axio Imager M2, Zeiss). To allow irradiation of the sample from the light source below the stage, the tank had a transparent window aligned below the sample, Fig. S1. A glass cover slide was used to seal the window. The slide was slightly tilted to prevent acoustic reflections from entering the chamber. The transducer was immersed in the tank such that its face was perpendicular to the fluid channel in the test chamber. The chamber was positioned so that the fluid channel-gel boundary falls within the focal zone. A 60$\times$ water dipping lens (W N-Achroplan 63$\times$/0.9) was used to visualize the bubble interactions at the boundary. The optical focus was aligned with the transducer focus using the hydrophone. A high-speed camera was also mounted on the microscope using a double connector. The camera allowed for recordings at 44,000 frames per second with a field of view of 85 \textmu m$\times$58 \textmu m. Two silicone tubes were placed in the inlet and outlet ports to replace the microbubble solution with a fresh bolus for every experiment. The tubes were attached to a syringe pump via two syringes to maintain a controlled flow rate during solution replacement.

\subsection*{Experimental procedure}
After a fresh bolus of bubbles was introduced in the fluid channel, an ultrasound pulse of 1 ms duration and 150 Hz pulse repetition frequency was applied at 1 MHz. These parameters fall within the range used for drug delivery \cite{petit2015fibrin, coussios2008applications}. The acoustic pressure at the focus was varied from 0.3 MPa to 0.5 MPa. The highest range was chosen to avoid inertial cavitation \cite{lin2017sonovuecavitation}. The high-speed camera was triggered by the same function generator that was used to trigger the transducer. Imaging began 100 \textmu s before the pulse arrived at the fluid-gel boundary. Each experiment used 8 pulses, determined from preliminary tests to be the maximum number required for bubbles to exit the imaging window for all fibrin densities and pressures. Imaging was performed throughout the ultrasound exposure. The location where bubble penetration occurred was identified using fiducial markers.   

\subsection*{High speed imaging analysis}

Bubble detection and tracking were done via a custom Python script. Bubbles and the edge/boundary of the hydrogel were detected as connected regions in the image using the skimage measure package. This detection method is robust under changing bubble radius. For the videos with relatively poor contrast (for a lower density of fibrin network), a custom script was written using OpenCV to detect the boundary by manually clicking on the images. Another custom script was then written to track the detected bubbles and boundary using the link function from the trackpy package \cite{crocker1996methodstrackpy}. The position data obtained from the code was post-analyzed to estimate the total distance traveled by the bubble along the x and y directions, and the instantaneous speed. Bubble speeds are calculated only when the ultrasound is on during the 1 ms period. Bubbles that penetrate more than 25 \textmu m into the mesh are considered for analysis. It is worth noting that bubbles were observed moving in and out of the focal plane within the imaging window due to their three-dimensional motion. This affects tracking and may lead to underestimation of the total distance covered during ultrasound pulses.  

\subsection*{Confocal Imaging}
Confocal imaging was performed to assess the pore size of fibrin networks. Fluorescent fibrin gels that were not exposed to ultrasound were imaged (Leica TCS SP8, Leica). Pore size was estimated by segmenting and thresholding a 2D image using ImageJ, followed by pore size estimation using the DiameterJ plugin.
The extent of nanobead penetration from bubble action was also qualitatively estimated on a subset of samples. Imaging was performed at the identified locations of bubble penetration (using fiducial markers) with a 60$\times$ objective lens. Z- stack images for a depth of 300 \textmu m were acquired at 10 \textmu m step size. These images were stitched to develop 3-D profiles of nanobead penetration along the paths created by the bubbles in the fibrin mesh. 

\subsection*{Uniaxial indentation measurements}
Fibrin gels were created in acrylic molds of 1.5$\times$1.5$\times$3.5 cm\textsuperscript{3}. Indentation experiments were carried out on a mechanical tester (5942, Instron) within 3 hours of gel formation. The bottom grippers of the tester were used to hold the mold containing the sample. An indenter with a 50 \textmu m radius was attached to the top rubber grippers that could move vertically at a desired rate. The chosen indenter diameter-to-sample dimension ratio was kept below 1:10 to ensure that edge effects do not influence the measurements \cite{hill1954mathematicalplasticity}. The tester was operated with a 1 kN load cell (2580-1KN, Instron). Before beginning the experiment, the indenter was placed right above the sample at a pre-load of 1 mN to ensure contact. During the experiment, the indenter was moved at a speed of 80 \textmu m/s until failure of the fibrin network was observed (4-5 mm maximum displacement). Throughout the experiment, the gels were immersed in PBS solution to prevent dehydration. Force-displacement curves were obtained for $n=3$ samples of each density, and the mean response is shown in Fig. 2A.

\subsection*{AFM measurements}
AFM is used to evaluate the stiffness and the role of repeated forcing on fibrin networks. The measurements were performed on the Bio-AFM (NanoWizard 4, JPK Instruments) using a colloidal probe (CP-FM-BSG-B-5, CP-CONT-BSG-B-5) with 10 \textmu m diameter glass bead. The probe/cantilever was first calibrated on a clean glass slide to measure the sensitivity and spring constant (using Thermal Noise method) \cite{afmhydrogelproto2021measuring}. The cantilever was then recalibrated in PBS using the same procedure to eliminate any effects of liquid on sensitivity and spring constant. After stabilization of the cantilever in liquid, fibrin hydrogels, formed on ozone-treated glass slides to improve adhesion, were mounted. The sample size was 1.5 cm$\times$1.5 cm$\times$1.5 mm. AFM measurements were performed in Force Spectroscopy mode. 

To measure the stiffness of the sample, the probe completed a single cycle of approach, indent, and retract phases to obtain a force-displacement (F-D) curve at a slow rate of 8 \textmu m/s. F-D curves were collected at 5 random locations, with each location comprising 5$\times$5 grid points. In total, 25 F-D curves were collected for each sample. A sample size of $n=3$ was taken for each density. We obtained the Young's modulus, $E$, using the established Oliver-Pharr model from the force-displacement curves, to account for viscoelastic effects in a semi-infinite medium \cite{afmhydrogelproto2021measuring}:

\begin{align*}
E&=\frac{\sqrt{\pi}(1-\nu^2)S}{2\sqrt{A}}
\end{align*}

where $\nu$ is the Poisson's ratio (0.5 in our case) and $S$ is the contact stiffness calculated as $S=dP/dh$. This expression corresponds to the slope of the upper portion of the unloading F-D curve where $P$ is the load at displacement $h$. $A$ is the contact area calculated as 

\begin{align*}
A&=\pi(2R_{tip}h_{c}-h_c^2)
\end{align*}

where $R_{tip}$ is the tip radius and $h_c=h_{max}-h_s$. The variable $h_{max}$ is the maximum indentation depth and $h_S=\epsilon P_{max}/S$ where $P_{max}$ is the maximum peak load. $\epsilon$ is 0.75 for spherical indenters. The mean and the distribution of stiffness values are shown in Fig. 1A across all fiber densities.

To analyze the role of repeated loading on fibrin networks, indentations were performed at a rate of 80 \textmu m/s at a single location. The maximum possible displacement was set to 15 \textmu m, and the peak applied force was varied between 50-500 nN to examine the role of force and corresponding stress magnitude on fatigue response. The experiment was performed for a maximum of 4000 cycles or until the indenter position reached 15 \textmu m at a given location. The collected F-D curves without a clear contact point, high noise due to mechanical interference, or large negative stress were discarded \cite{afmhydrogelproto2021measuring}.  Each F-D curve was post-processed to extract the modulus, hysteresis, and deformation. The change in modulus, Figs. 3D and 3E, was estimated as the difference in modulus between the first and last cycle. The deformation was calculated as the difference between the contact point of the $n^{th}$ curve and the first F-D curve during loading, Figs. 3F and 3G. The hysteresis is the area enclosed within the loading-unloading curve of a cycle. The change in hysteresis, Figs. 3H and 3I, was calculated as the difference in hysteresis between the first and last cycle. While sample size varied as a few F-D curves were discarded in post-analysis, a sample size of at least $n\geq 3$ was maintained. 

\subsection*{Modeling of bubble radial dynamics}
The radial dynamics of the bubble was estimated using the established Marmottant model or modified Rayleigh-Plesset equation \cite{marmottant2005model}. Briefly, for a bubble of equilibrium radius $R_0$ oscillating in an acoustic field of pressure $P_{i}$ and frequency $\omega$, the radius, $R$, evolution in time domain is estimated from:

\begin{align*}
\rho_0(R\ddot{R}+\frac{3}{2}\dot{R}^2)&=(1-\frac{3\gamma\dot{R}}{c_0})(P_0+\frac{2\sigma(R_0)}{R_0})(\frac{R_0}{R})^{3\gamma}-\frac{2\sigma (R)}{R}\\
&-\frac{4\dot{R}}{R}(\mu_0+\frac{\kappa_s}{R})-P_0+P_i\sin{\omega t}
\end{align*}

where $\rho_0$, $\gamma$, $c_0$, $\mu_0$, and $P_0$ are density, polytropic exponent, speed of sound, dynamic viscosity, and ambient pressure of the medium, respectively. The overdot and double overdot represents the first and second time derivative of a quantity, respectively. The shell with dilational viscosity, $\kappa_s$ has radius-dependent surface tension, $\sigma$ expressed as

\[
\sigma(R)=\;\left\{
\begin{array}{l@{\qquad}l}
0 & R \le R_{\mathrm{buck}},\\[4pt]
\chi\!\left[\left(\dfrac{R}{R_{\mathrm{buck}}}\right)^{\!2} - 1\right]
  & R_{\mathrm{buck}} \le R \le R_{\mathrm{rupt}},\\[6pt]
\sigma_0 & R \ge R_{\mathrm{rupt}}
\end{array}
\right.
\]

where $\chi$ is the elastic compression modulus of the shell, $\sigma_0$ is the surface tension of the fluid, $R_{buck}$ is the buckling radius with $R_{buck}=R_0$, and $R_{rupt}$ is the rupture radius defined as $R_{rupt}=R_{buck}\sqrt{1+\sigma_0/\chi}$. The values used in our study to estimate bubble radial oscillation are $\rho_0$ = 997.8 kg/m\textsuperscript{3}, $\gamma$ = 1.098, $c_o$ = 1481 m/s, $\mu_0$ = 9.5x10\textsuperscript{-4} Pas, $\kappa_s$ = 3.2 x 10\textsuperscript{-9}, $\chi$ = 0.3, $\sigma_0$ = 0.0725. 

The dominant forces from the bubble exerted in the surrounding medium are the primary radiation force, $F_{rad}$ and the force from the radial oscillation of the bubble, $F_{osc}$. Secondary forces, such as the acoustic streaming force, are neglected as they are 1-2 orders below primary forces \cite{elder1959cavitationstreaming}. 

\begin{align*}
F_{rad}&=-V(t)\frac{\partial P_{ac}(t)}{\partial x}
\end{align*}
\begin{align*}
F_{osc}=\rho_02\pi R(t)^2(R\ddot{R}+\dot{R(t)^2})
\end{align*}

where $V(t)$ is the volume of the bubble. The temporal evolution of bubble radius and corresponding forces was estimated for experimental conditions in Fig. 1 using this model.

\subsection*{Statistical analysis}
To compare the variation of bubble speed and distance traveled across densities and ultrasound pressures, one-way ANOVA with Tukey’s honest significant criterion was performed to compare multiple groups, Fig. 1. Similarly, the influence of density and maximum allowed stress on hysteresis, Young's modulus, and deformation was also compared, Fig. 2. The statistical difference between groups was considered significant for $p<0.05$. All statistical analyses were performed in MATLAB 2025b (MathWorks).


\section*{Acknowledgements}

This work was supported by Humboldt Research Fellowship from the Alexander von Humboldt Foundation, Max Planck Society, and European Research Council (ERC) Advanced Grant SoMMoR project with grant no. 834531.

\section*{Competing interests}

The authors declare no competing interests.

\section*{Author contributions}

A.B designed research; A.B performed research; A.B and G.G. analyzed data; A.B. wrote the paper, and A.B., G.G., and M.S edited the paper.

\section*{Data availability}

The data that support the findings of this study are available from
the corresponding authors upon reasonable request.



\begin{thebibliography}{99}
 
\bibitem{leighton2012acoustic}
Leighton,~T.
\newblock {\em The acoustic bubble}.
\newblock Academic press, 2012.
 
\bibitem{burgess2015focusedbbb}
Burgess,~A., Shah,~K., Hough,~O., and Hynynen,~K.
\newblock Focused ultrasound-mediated drug delivery through the blood--brain barrier.
\newblock {\em Expert review of neurotherapeutics}, 15(5), 477--491, 2015.
 
\bibitem{arrieta2024nueromod}
Arrieta,~V. et~al.
\newblock Ultrasound-mediated delivery of doxorubicin to the brain results in immune modulation and improved responses to PD-1 blockade in gliomas.
\newblock {\em Nature Communications}, 15(1), 4698, 2024.
 
\bibitem{miller2002sonoporation}
Miller,~D., Pislaru,~S., and Greenleaf,~J.
\newblock Sonoporation: mechanical DNA delivery by ultrasonic cavitation.
\newblock {\em Somatic cell and molecular genetics}, 27(1), 115--134, 2002.
 
\bibitem{wright2012vitrothrombo}
Wright,~C., Hynynen,~K., and Goertz,~D.
\newblock In vitro and in vivo high-intensity focused ultrasound thrombolysis.
\newblock {\em Investigative radiology}, 47(4), 217--225, 2012.
 
\bibitem{khokhlova2015histotripsyrev}
Khokhlova,~V. et~al.
\newblock Histotripsy methods in mechanical disintegration of tissue: Towards clinical applications.
\newblock {\em International journal of hyperthermia}, 31(2), 145--162, 2015.
 
\bibitem{coussios2008applications}
Coussios,~C., and Roy,~R.
\newblock Applications of acoustics and cavitation to noninvasive therapy and drug delivery.
\newblock {\em Annu. Rev. Fluid Mech.}, 40(1), 395--420, 2008.
 
\bibitem{datta2006correlationthrombo}
Datta,~S. et~al.
\newblock Correlation of cavitation with ultrasound enhancement of thrombolysis.
\newblock {\em Ultrasound in medicine \& biology}, 32(8), 1257--1267, 2006.
 
\bibitem{epidemiologyvte2003circ}
White,~R.
\newblock The epidemiology of venous thromboembolism.
\newblock {\em Circulation}, 107(23\_suppl\_1), I--4, 2003.
 
\bibitem{dvtrisk2014dvt}
Alper,~E. et~al.
\newblock Risk stratification model: lower-extremity ultrasonography for hospitalized patients with suspected deep vein thrombosis.
\newblock {\em Journal of General Internal Medicine}, 33, 21--25, 2018.
 
\bibitem{heartandstroke2017heart}
Benjamin,~E. et~al.
\newblock Heart disease and stroke statistics—2017 update: a report from the American Heart Association.
\newblock {\em circulation}, 135(10), e146--e603, 2017.
 
\bibitem{weisel2011coronary}
Silvain,~J. et~al.
\newblock Composition of coronary thrombus in acute myocardial infarction.
\newblock {\em Journal of the American College of Cardiology}, 57(12), 1359--1367, 2011.
 
\bibitem{thrombolysistime2023thrombuscomp}
Ho-Tin-No{\'e},~B., Desilles,~J., and Mazighi,~M.
\newblock Thrombus composition and thrombolysis resistance in stroke.
\newblock {\em Research and Practice in Thrombosis and Haemostasis}, 7(4), 100178, 2023.
 
\bibitem{petit2015fibrin}
Petit,~B. et~al.
\newblock Fibrin degradation during sonothrombolysis--Effect of ultrasound, microbubbles and tissue plasminogen activator.
\newblock {\em Journal of Drug Delivery Science and Technology}, 25, 29--35, 2015.
 
\bibitem{sutton2013clot}
Sutton,~J., Ivancevich,~N., Perrin  Jr,~S., Vela,~D., and Holland,~C.
\newblock Clot retraction affects the extent of ultrasound-enhanced thrombolysis in an ex vivo porcine thrombosis model.
\newblock {\em Ultrasound in medicine \& biology}, 39(5), 813--824, 2013.
 
\bibitem{hendleydoublingdown2022plos}
Hendley,~S. et~al.
\newblock (More than) doubling down: Effective fibrinolysis at a reduced rt-PA dose for catheter-directed thrombolysis combined with histotripsy.
\newblock {\em Plos one}, 17(1), e0261567, 2022.
 
\bibitem{holland2024histotripsyholl}
Yang,~S., Zemzemi,~C., Escudero,~D., Vela,~D., Haworth,~K., and Holland,~C.
\newblock Histotripsy and Catheter-Directed Lytic: Efficacy in Highly Retracted Porcine Clots In Vitro.
\newblock {\em Ultrasound in Medicine \& Biology}, 50(8), 1167--1177, 2024.
 
\bibitem{fibrinfunction2018cardth}
Baker,~S., and Ariëns,~R.
\newblock Fibrin clot structure and function: a novel risk factor for arterial and venous thrombosis and thromboembolism.
\newblock {\em Cardiovascular thrombus. Amsterdam: Elsevier}, 2018.
 
\bibitem{rmbolism2021artscl}
Alkarithi,~G., Duval,~C., Shi,~Y., Macrae,~F., and Ariëns,~R.
\newblock Thrombus structural composition in cardiovascular disease.
\newblock {\em Arteriosclerosis, thrombosis, and vascular biology}, 41(9), 2370--2383, 2021.
 
\bibitem{fibrinextraordextensi2006weisel}
Liu,~W. et~al.
\newblock Fibrin fibers have extraordinary extensibility and elasticity.
\newblock {\em Science}, 313(5787), 634--634, 2006.
 
\bibitem{brown2009multiscale}
Brown,~A., Litvinov,~R., Discher,~D., Purohit,~P., and Weisel,~J.
\newblock Multiscale mechanics of fibrin polymer: gel stretching with protein unfolding and loss of water.
\newblock {\em science}, 325(5941), 741--744, 2009.
 
\bibitem{alteredfibthrombolysis2002circ}
Mills,~J., Ariëns,~R., Mansfield,~M., and Grant,~P.
\newblock Altered fibrin clot structure in the healthy relatives of patients with premature coronary artery disease.
\newblock {\em Circulation}, 106(15), 1938--1942, 2002.
 
\bibitem{fibronolysistight1996thrandhem}
Fatah,~K., Silveira,~A., Tornvall,~P., Karpe,~F., Blomb{\"a}ck,~M., and Hamsten,~A.
\newblock Proneness to formation of tight and rigid fibrin gel structures in men with myocardial infarction at a young age.
\newblock {\em Thrombosis and haemostasis}, 76(10), 535--540, 1996.
 
\bibitem{wan2015cavitationgail}
Wan,~M., Feng,~Y., and Ter  Haar,~G.
\newblock Cavitation in biomedicine.
\newblock {\em Netherlands: Springer}, 2015.
 
\bibitem{bezer2025microbubblebrain}
Bezer,~J. et~al.
\newblock Microbubble dynamics in brain microvessels.
\newblock {\em PloS one}, 20(2), e0310425, 2025.
 
\bibitem{ferrara2009microbubble}
Caskey,~C., Qin,~S., Dayton,~P., and Ferrara,~K.
\newblock Microbubble tunneling in gel phantoms.
\newblock {\em The Journal of the Acoustical Society of America}, 125(5), EL183--EL189, 2009.
 
\bibitem{acconcia2013interactions}
Acconcia,~C., Leung,~B., Hynynen,~K., and Goertz,~D.
\newblock Interactions between ultrasound stimulated microbubbles and fibrin clots.
\newblock {\em Applied Physics Letters}, 103(5), 2013.
 
\bibitem{retractedcompos2018jtt}
Zabczyk,~M., Natorska,~J., and Undas,~A.
\newblock Erythrocyte compression index is impaired in patients with residual vein obstruction.
\newblock {\em Journal of Thrombosis and Thrombolysis}, 46, 31--38, 2018.
 
\bibitem{zkabczyk2023porosityclots}
Ząbczyk,~M., Ariëns,~R., and Undas,~A.
\newblock Fibrin clot properties in cardiovascular disease: from basic mechanisms to clinical practice.
\newblock {\em Cardiovascular research}, 119(1), 94--111, 2023.
 
\bibitem{afmhydrogelproto2021measuring}
Norman,~M., Ferreira,~S., Jowett,~G., Bozec,~L., and Gentleman,~E.
\newblock Measuring the elastic modulus of soft culture surfaces and three-dimensional hydrogels using atomic force microscopy.
\newblock {\em Nature Protocols}, 16(5), 2418--2449, 2021.
 
\bibitem{wufsus2013clotpermeability}
Wufsus,~A., Macera,~N., and Neeves,~K.
\newblock The hydraulic permeability of blood clots as a function of fibrin and platelet density.
\newblock {\em Biophysical journal}, 104(8), 1812--1823, 2013.
 
\bibitem{shibeko2020redistributiontpaflux}
Shibeko,~A., Chopard,~B., Hoekstra,~A., and Panteleev,~M.
\newblock Redistribution of TPA fluxes in the presence of PAI-1 regulates spatial thrombolysis.
\newblock {\em Biophysical journal}, 119(3), 638--651, 2020.
 
\bibitem{mican2019structuralproteinsizedrug}
Mican,~J., Toul,~M., Bednar,~D., and Damborsky,~J.
\newblock Structural biology and protein engineering of thrombolytics.
\newblock {\em Computational and structural biotechnology journal}, 17, 917--938, 2019.
 
\bibitem{boots2022sawtoothrupture}
Boots,~J. et~al.
\newblock Quantifying bond rupture during indentation fracture of soft polymer networks using molecular mechanophores.
\newblock {\em Physical Review Materials}, 6(2), 025605, 2022.
 
\bibitem{montanari2023puncturingrupture2}
Montanari,~M., Brighenti,~R., Terzano,~M., and Spagnoli,~A.
\newblock Puncturing of soft tissues: experimental and fracture mechanics-based study.
\newblock {\em Soft Matter}, 19(20), 3629--3639, 2023.
 
\bibitem{piechocka2010fibrinhierarchy}
Piechocka,~I., Bacabac,~R., Potters,~M., MacKintosh,~F., and Koenderink,~G.
\newblock Structural hierarchy governs fibrin gel mechanics.
\newblock {\em Biophysical journal}, 98(10), 2281--2289, 2010.
 
\bibitem{marmottant2005model}
Marmottant,~P. et~al.
\newblock A model for large amplitude oscillations of coated bubbles accounting for buckling and rupture.
\newblock {\em The Journal of the Acoustical Society of America}, 118(6), 3499--3505, 2005.
 
\bibitem{lee1993acousticrad}
Lee,~C., and Wang,~T.
\newblock Acoustic radiation force on a bubble.
\newblock {\em The Journal of the Acoustical Society of America}, 93(3), 1637--1640, 1993.
 
\bibitem{jimenez2005bubbleosci}
Jim{\'e}nez-Fern{\'a}ndez,~J., and Crespo,~A.
\newblock Bubble oscillation and inertial cavitation in viscoelastic fluids.
\newblock {\em Ultrasonics}, 43(8), 643--651, 2005.
 
\bibitem{elder1959cavitationstreaming}
Elder,~S.
\newblock Cavitation microstreaming.
\newblock {\em The Journal of the Acoustical Society of America}, 31(1), 54--64, 1959.
 
\bibitem{liu2024fibrinfatigueandvariablerateloading}
Liu,~S., Bahmani,~A., Ghezelbash,~F., and Li,~J.
\newblock Fibrin clot fracture under cyclic fatigue and variable rate loading.
\newblock {\em Acta Biomaterialia}, 177, 265--277, 2024.
 
\bibitem{mooney2006indentationrate}
Mooney,~R. et~al.
\newblock Indentation micromechanics of three-dimensional fibrin/collagen biomaterial scaffolds.
\newblock {\em Journal of materials research}, 21(8), 2023--2034, 2006.
 
\bibitem{litvinovstrengthfibrin}
Litvinov,~R., Tutwiler,~V., Maksudov,~F., Barsegov,~V., and Weisel,~J.
\newblock Strength and deformability of fibrin clots: Biomechanics, thermodynamics, and mechanisms of rupture.
 
\bibitem{kliuchnikov2025strengthdamage}
Kliuchnikov,~E. et~al.
\newblock Strength, deformability, damage and fracture toughness of fibrous material networks: Application to fibrin clots.
\newblock {\em Acta Biomaterialia}, 201, 347--359, 2025.
 
\bibitem{vlaisavljevich2014effectsstiffnesshisto}
Vlaisavljevich,~E., Kim,~Y., Owens,~G., Roberts,~W., Cain,~C., and Xu,~Z.
\newblock Effects of tissue mechanical properties on susceptibility to histotripsy-induced tissue damage.
\newblock {\em Physics in medicine and biology}, 59(2), 253--270, 2014.
 
\bibitem{Highspeeddeformationinertial2023franck}
McGhee,~A. et~al.
\newblock High-speed, full-field deformation measurements near inertial microcavitation bubbles inside viscoelastic hydrogels.
\newblock {\em Experimental Mechanics}, 63(1), 63--78, 2023.
 
\bibitem{bai2019fatiguehydrogel}
Bai,~R., Yang,~J., and Suo,~Z.
\newblock Fatigue of hydrogels.
\newblock {\em European Journal of Mechanics-A/Solids}, 74, 337--370, 2019.
 
\bibitem{liu2025molecularshakedown}
Liu,~D., Jia,~Z., Liu,~B., Hou,~J., Bao,~R., and Wang,~M.
\newblock Molecular Creep Induced Fatigue Rupture of Fibrin Clots.
\newblock {\em Advanced Science}, 12(38), e05109, 2025.
 
\bibitem{chinh2008shakedown}
Chinh,~P.
\newblock On shakedown theory for elastic--plastic materials and extensions.
\newblock {\em Journal of the Mechanics and Physics of Solids}, 56(5), 1905--1915, 2008.
 
\bibitem{liu2010mechanicalgutholdsiglefibers}
Liu,~W., Carlisle,~C., Sparks,~E., and Guthold,~M.
\newblock The mechanical properties of single fibrin fibers.
\newblock {\em Journal of thrombosis and haemostasis}, 8(5), 1030--1036, 2010.
 
\bibitem{litvinov2017fibrinorigin}
Litvinov,~R., and Weisel,~J.
\newblock Fibrin mechanical properties and their structural origins.
\newblock {\em Matrix Biology}, 60, 110--123, 2017.
 
\bibitem{ramanujam2023biomechanicsenergeticsfibrin}
Ramanujam,~R. et~al.
\newblock Biomechanics, energetics, and structural basis of rupture of fibrin networks.
\newblock {\em Advanced Healthcare Materials}, 12(27), 2300096, 2023.
 
\bibitem{jones2006trappingstride}
Jones,~P., Stride,~E., and Saffari,~N.
\newblock Trapping and manipulation of microscopic bubbles with a scanning optical tweezer.
\newblock {\em Applied physics letters}, 89(8), 2006.
 
\bibitem{beekers2022internalizationkooiman}
Beekers,~I. et~al.
\newblock Internalization of targeted microbubbles by endothelial cells and drug delivery by pores and tunnels.
\newblock {\em Journal of Controlled Release}, 347, 460--475, 2022.
 
\bibitem{postema2004ultrasound}
Postema,~M., Marmottant,~P., Lanc{\'e}e,~C., Hilgenfeldt,~S., and De  Jong,~N.
\newblock Ultrasound-induced microbubble coalescence.
\newblock {\em Ultrasound in medicine \& biology}, 30(10), 1337--1344, 2004.
 
\bibitem{brujan2001dynamicsjetting}
Brujan,~E., Nahen,~K., Schmidt,~P., and Vogel,~A.
\newblock Dynamics of laser-induced cavitation bubbles near elastic boundaries: influence of the elastic modulus.
\newblock {\em Journal of Fluid Mechanics}, 433, 283--314, 2001.
 
\bibitem{lin2017sonovuecavitation}
Lin,~Y. et~al.
\newblock Effect of acoustic parameters on the cavitation behavior of SonoVue microbubbles induced by pulsed ultrasound.
\newblock {\em Ultrasonics Sonochemistry}, 35, 176--184, 2017.
 
\bibitem{vince2023highspeedmicrospheretransport}
Vince,~J., Lewis,~A., and Stride,~E.
\newblock High-speed imaging of microsphere transport by cavitation activity in a tissue-mimicking phantom.
\newblock {\em Ultrasound in Medicine \& Biology}, 49(6), 1415--1421, 2023.
 
\bibitem{bertula2019strainstiffeiningagarose}
Bertula,~K., Martikainen,~L., Munne,~P., Hietala,~S., Klefstr{\"o}m,~J., and Ikkala,~O.
\newblock Strain-stiffening of agarose gels.
\newblock {\em Acs macro letters}, 8(6), 670--675, 2019.
 
\bibitem{munster2013strainhistory}
M{\"u}nster,~S., Jawerth,~L., Leslie,~B., Weitz,~J., Fabry,~B., and Weitz,~D.
\newblock Strain history dependence of the nonlinear stress response of fibrin and collagen networks.
\newblock {\em Proceedings of the National Academy of Sciences}, 110(30), 12197--12202, 2013.
 
\bibitem{chawla2024henalcollagenfatigue}
Chawla,~D., Thao,~A., Eriten,~M., and Henak,~C.
\newblock Articular cartilage fatigue causes frequency-dependent softening and crack extension.
\newblock {\em Journal of the Mechanical Behavior of Biomedical Materials}, 160, 106753, 2024.
 
\bibitem{vazquez2019cartilagefatigue}
Vazquez,~K., Andreae,~J., and Henak,~C.
\newblock Cartilage-on-cartilage cyclic loading induces mechanical and structural damage.
\newblock {\em Journal of the mechanical behavior of biomedical materials}, 98, 262--267, 2019.
 
\bibitem{sennoga2012sonovuesize}
Sennoga,~C. et~al.
\newblock Evaluation of methods for sizing and counting of ultrasound contrast agents.
\newblock {\em Ultrasound in medicine \& biology}, 38(5), 834--845, 2012.
 
\bibitem{crocker1996methodstrackpy}
Crocker,~J., and Grier,~D.
\newblock Methods of digital video microscopy for colloidal studies.
\newblock {\em Journal of colloid and interface science}, 179(1), 298--310, 1996.
 
\bibitem{hill1954mathematicalplasticity}
Hill,~R.
\newblock The Mathematical Theory of Plasticity: Oxford (1950).
\newblock {\em Johnson, KL: Proc. Roy. Soc. A}, 230, 531, 1954.
 
\end{thebibliography}







\end{document}